\documentclass[twocolumn,showpacs,preprintnumbers,amsmath,amssymb]{revtex4}
\usepackage{graphicx}
\usepackage{amsmath}

\begin{document}

\title{ \bf Exact wave-packet decoherence dynamics
in a discrete spectrum environment}

\author{ \bf Matisse W. Y. Tu and Wei-Min Zhang}
\email{wzhang@mail.ncku.edu.tw} \affiliation{Department of
Physics, National Cheng Kung University, Tainan
70101, Taiwan \\
National Center for Theoretical Science, Tainan 70101, Taiwan}

\begin{abstract}
We find an exact analytical solution of the reduced density matrix
from the Feynman-Vernon influence functional theory for a wave
packet influenced by an environment containing a few discrete
modes. We obtain two intrinsic energy scales relating to the time
scales of the system and the environment. Different relationship
between these two scales alters the overall form of the solution
 of the system. We also introduce a decoherence measure
for a single wave packet which is defined as the ratio of
Schr\"{o}dinger uncertainty over the delocalization extension of
the wave packet and characterizes the time-evolution behavior of
the off-diagonal reduced density matrix element. We utilize the
exact solution and the docherence measure to study the wave packet
decoherence dynamics. We further demonstrate how the dynamical
diffusion of the wave packet leads to non-Markovian decoherence in
such a microscopic environment.
\end{abstract}

\pacs{03.65.Yz, 03.65.Db, 42.50.Lc}

\date{September 25, 2008}

\maketitle

\section{INTRODUCTION}
Wave packet dynamics has been extensively studied since the birth
of quantum mechanics \cite{sch26}.  Studying decoherence dynamics
of a wave packet still attracts attention in various research
topics, from quantum dissipation dynamics, quantum cosmology,
quantum measurement theory, quantum information processing, to the
foundation of quantum mechanics \cite{zuk}. The process of quantum
decoherence is triggered by the interaction of the system with its
environment. A widely used model of the environment is a set of
harmonic oscillators \cite{fey,cal}. Environments with continuous
distribution of oscillator frequencies have been extensively
studied concerning the dissipation effects and the induced
decoherence
\cite{zuk1,clann,fbq,hu,hak,ez,zur,jpz,braun,str,kazu,aw,an,Chou07}.
We shall consider in this paper an environment containing finite
discrete modes.

Wave packet decoherence induced by an environment containing a few
discrete modes has received attention in recent years in the study
of quantum information processing using wave packets. There are
many situations where the environment has a discrete spectrum. The
decoherence of molecule vibrational wave packet due to the
vibrational-rotational coupling is a typical example
\cite{brf,chc}.  Other physical situations where environments have
only a few modes being relevant include electromagnetic field in
cavity QED, molecular qubit decoherence in glassy environments
\cite{wong}, and electron transfer and exciton decoherence
dynamics in proteins \cite{et}, to name a few. In fact, many
environments associating with manipulations and measurements in
quantum information processing involve a finite and discrete
modes. The decoherence process under such circumstances must be
studied with the back-reaction of the environment to the system
being fully taken into account. The result is expected to be
different from that happens with a continuous spectrum
environment.

For making comparison between the mechanisms of decoherence in a
few discrete modes and a continuous spectrum environment, we take
the Caldeira-Leggett model \cite{cal} with a discrete spectrum
environment as a representative one. In this model, the principal
system is a particle in a harmonic trap and the environment is a
set of noninteracting harmonic oscillators each of which linearly
coupled to the principal system in their coordinates. Many people
have studied this model, some have done research on models
slightly modified from this concerning topics not only of
decoherence but many other issues of open quantum
systems\cite{zuk1,clann,legt,fbq,hu,zur,jpz,str,kazu,braun,aw,an,Chou07,hak,gr,ez,crm,wss,breuer}.
An exact master equation to the motion of the central oscillator
with a general spectral density at an arbitrary temperature was
given in \cite{hu} and derived alternatively in \cite{jh}. The
exact master equation lies on the dissipation dynamical equation
which fully takes into account the back-action effects of the
environment on the system. Here we present an analytical solution
to the dissipation dynamical equation to directly describe the
non-Markovian processes of the central particle. And the wave
packet dynamics can be expressed in terms of the solution to the
dissipation dynamical equation without invoking the master
equation. As a result, a qualitative change of the dynamics of the
central particle is directly read from the exact solution to the
dissipation dynamical equation we obtained. The resulting effect
can be used to analyze decoherence and instability of the system
under the influence of the environment.

The paper is organized as follows. In the next section, we briefly
describe the Feynman-Vernon influence functional approach to a
quantum harmonic oscillator interacting linearly with a general
environment of $N$ modes. We then in the following section solve
analytically the reduced dynamics of the system. We obtain two
intrinsic energy scales relating to the time scales of the system
and the environment. We will explicitly show how the different
energy scales between the system and the environment alters
qualitatively the dynamics of the system. We also take the
continuum limit of the environmental frequencies to the Ohmic
spectrum\cite{legt} and recover the solution previously obtained
in \cite{cal,jpz}. In section IV we study the evolution of a
single wave packet.  We introduce a decoherence measure for the
single wave packet evolution. Non-Markovian decoherence of the
wave packet in such an environment with a few discrete modes is
demonstrated by a modified Drude's spectral density \cite{mdru}.
Physical implications of our results will be given and discussed
in the conclusion section. The detailed mathematics is left in the
Appendix.

\section{Time evolution of open quantum systems}
Following many works on Caldeira-Leggett model \cite{fey,cal}, we
briefly review the main results obtained previously that will be
used later in this work. The Hamiltonian of a particle in a
harmonic trap linearly coupled with an environment is given by
\begin{align}
H =\Big(\frac{P^2}{2M}+\frac{M\Omega^2X^2}{2}\Big)
+&\sum_{j=1}^{N}\Big(\frac{p_j^2}{2m}+\frac{m\omega_j^2q_j^2}{2}
\Big)\nonumber\\+&\sum_{j=1}^{N}C_jXq_j \label{hamil}.
\end{align}
The first two terms in the bracket is the Hamiltonian of the
system and the second term as a summation is that of the
environment consisting of a finite $N$ discrete modes. The third
term is the interaction Hamiltonian between the system and the
environment. The notations follow the convention. The requirement
of a finite $N$ discrete modes in the environment makes our
consideration different from original Caldeira-Leggett model. It
should be particularly noticed that in Caldeira-Leggett model an
additional counter-term, $\sum_{i=1}^N{C_i^2\over
2m\omega_i^2}X^2$, is usually added to Eq.~(\ref{hamil}) in order
to study the generic behavior of dissipation dynamics for a
harmonic oscillator. Such a counter-term exactly cancels the
coupling-induced frequency-renormalization effect \cite{clann}.
Since our interest lies on the environment with a finite discrete
modes where no high frequency cut-off is introduced so that no
counter-term is needed for Eq.~(\ref{hamil}), according to the
standard renormalization theory \cite{rent}. As it has also been
pointed out \cite{clann,clann1} the coupling-induced
frequency-renormalization effect in many cases is a physical
observable effect that has to be taken into account. In fact, the
study of quantum decoherence should fully explore the environment
induced dissipation, fluctuation as well as renormalization
effects on the system.  Nonexistence of the counter-term for the
present case of a discrete spectrum environment enables us to
explore the decoherence dynamics of wave packets purely induced by
the interaction with the environment without ambiguity.

By using the common assumption that the system and the environment
is initially uncorrelated while the environment starts with an
equilibrium state, the reduced density matrix of the system
evaluated in the position basis at time $t$ is given by
\cite{fey,cal}
\begin{eqnarray}
&& <x|\rho_{A}(t)|x'> \equiv \rho_{A}(x,x',t) \nonumber \\
&&~~~~~= \int dx_0dx_0'J(x,x',t|x_0,x_0',0)\rho_{A}(x_0,x_0',0),
 \end{eqnarray}
where $J(x,x',t|x_0,x_0',0)$ is the propagator which includes the
back-action effect of the environment on the system  and is given
by
\begin{eqnarray}
J(x,x',t|x_0,x_0',0)&=&\int \mathcal{D}[\textsl{x}]\mathcal{D}
[\textsl{x}']e^{ {i\over \hbar}
(S_A[\textsl{x}]-S_A[\textsl{x}'])}
\mathcal{F}[\textsl{x},\textsl{x}'],\nonumber\\ \label{prop0}
\end{eqnarray}
in which $S_A$ is the action of the system
\begin{equation}
S_A[\textsl{x}]=\int_0^t
d\tau\Big\{\frac{M\dot{\textsl{x}}^2(\tau)}{2}
-\frac{M\Omega^2\textsl{x}^2(\tau)}{2}\Big\},
\end{equation} and
$\mathcal{F}[\textsl{x},\textsl{x}']$ is the influence functional
describing the influence of the environment on the system. The
explicit form of the influence functional has been well solved
\cite{gr}:
\begin{widetext}
\begin{eqnarray}
\mathcal{F}[\textsl{x},\textsl{x}']&=& e^{{i\over \hbar} \big[
-\int_0^td\tau \int_0^\tau
d\tau'[\textsl{x}(\tau)-\textsl{x}'(\tau)]
K_I(\tau-\tau')[\textsl{x}(\tau')+\textsl{x}'(\tau')]
+i\int_0^td\tau \int_0^\tau
d\tau'[\textsl{x}(\tau)-\textsl{x}'(\tau)]
 K_R(\tau-\tau')[\textsl{x}(\tau')-\textsl{x}'(\tau')]\big]}, \label{if}
\end{eqnarray}
\end{widetext} where
\begin{equation}
K_I(\tau-\tau')=-\sum_{k=1}^{N}\frac{C_k^2}{2m\omega_k}\sin\omega_k(\tau-\tau')
\end{equation}
 and
\begin{equation} K_R(\tau-\tau')=\sum_{k=1}^{N}\frac{C_k^2}{2m\omega_k}
\coth{\frac{\hbar\omega_k\beta}{2}}\cos\omega_k(\tau-\tau'),
\end{equation} are defined as the dissipation and fluctuation Kernels
respectively, and $\beta$ is reciprocal of the product of the
initial equilibrium temperature and Boltzmann constant.

Eqs.~(\ref{prop0}-\ref{if}) shows that the effective action has a
quadratic form. Hence the path integral can be exactly carried out
with the stationary path method \cite{fey1}. By introducing the
new variables $
\texttt{R}(\tau)\equiv\frac{\textsl{x}(\tau)+\textsl{x}'(\tau)}{2}$
and $ \texttt{r}(\tau)\equiv \textsl{x}(\tau)-\textsl{x}'(\tau),$
we rewrite the propagator $J(x,x',t|x_0,x_0',0)$ as
$J(R,r,t|R_0,r_0,0)$.  The resulting propagator becomes
\cite{kazu,kzc}
\begin{widetext}
\begin{eqnarray}
J(R,r,t|R_0,r_0,0)&=& \tilde{N}_0 e^{\big\{ \frac{i}{\hbar}
M[r_0R_0\dot{u}_2(t)-r_0R\dot{u}_2(0)+R_0r\dot{u}_1(t)
-rR\dot{u}_1(0)] +i[\chi_{11}(t)r_0^2+(\chi_{12}+\chi_{21})(t)r_0r
+\chi_{22}(t)r^2]\big\}} , \label{prop}
\end{eqnarray}
\end{widetext}
 where $\chi_{ij}(t)$ ($i, j=1,2$) is defined by
\begin{equation} \label{fluctu1}
\chi_{ij}(t)=\int_0^td\tau \int_0^\tau
d\tau'v_i(\tau)K_R(\tau-\tau')v_j(\tau')
\end{equation}
 and $u_{i}$ and $v_{i}$ satisfy the
following dissipation dynamical equations of motion,
\begin{eqnarray}
\ddot{u}_i(\tau)+\Omega^2u_i(\tau)+{2\over M}\int_{0}^\tau
d\tau'K_I(\tau-\tau'){u}_i(\tau')=0,
\label{dissip1} \\
\ddot{{v}}_i(\tau)+\Omega^2{v}_i(\tau)-{2\over M}\int_{\tau}^t
d\tau'K_I(\tau-\tau'){v}_i(\tau')=0,
 \label{dissip2}
 \end{eqnarray}
subject to the boundary conditions $
v_1(0)=u_1(0)=v_2(t)=u_2(t)=1, v_1(t)=u_1(t)=v_2(0)=u_2(0)=0$.
$\tilde{N}_0$ is the contribution of fluctuations around the
stationary pathes which is independent of the position variables
for such a quadratic form of the Hamiltonian.

\section{Analytical Solution to the dissipation dynamics}
Since Eq.~(\ref{dissip2}) is the backward version of the
forwarding equation Eq.~(\ref{dissip1}), the propagator of the
system, Eq.~(\ref{prop}), is completely determined by solving the
dissipation dynamical equation (\ref{dissip1}).  The first two
terms in Eq.~(\ref{dissip1}) correspond to the classical equation
of motion for the coordinate of the central oscillator and the
third term with the dissipation kernel $K_{I}$ accounts for the
non-Markovian (memory) effect of the back-action of the
environment on the system. While, the fluctuation contribution
from the environment is given by Eq.~(\ref{fluctu1}).

On the other hand, the propagator Eq.~(\ref{prop}) is built upon
the solutions of the dissipation dynamical equations
(\ref{dissip1}-\ref{dissip2}). The solution of these equations
gives rise to the full dynamics of the system. Using the fact that
Eq.~(\ref{dissip2}) is the backward version of
Eq.~(\ref{dissip1}), $v_1(\tau)$ and $v_2(\tau)$ are related to
$u_1(\tau)$ and $u_2(\tau)$ by
$v_1(\tau)=u_2(t-\tau),v_2(\tau)=u_1(t-\tau)$. Furthermore,
$u_1(\tau)$ and $u_2(\tau)$ can be expressed as
$u_1(\tau)=\dot{Z}(\tau)-\frac{Z(\tau)}{Z(t)}\dot{Z}(t)$ and
$u_2(\tau)=\frac{Z(\tau)}{Z(t)}$ where $Z(\tau)$ obeys the same
equation as Eq.~(\ref{dissip1}),
\begin{eqnarray}
\ddot{Z}(\tau)+\Omega^2Z(\tau)+{2\over M}\int_{0}^\tau
d\tau'K_I(\tau-\tau'){Z}(\tau')=0,  \label{dissipz}
\end{eqnarray}
with the boundary conditions $Z(0)=0$ and $\dot{Z}(0)=1$.
 The solution of $Z(\tau)$ determines the
behaviors of $u_1(\tau),u_2(\tau)$ and $v_1(\tau),v_2(\tau)$ and
all the relevant dynamics of the reduced system. $Z(\tau)$ was
solved in cases of continuous spectra of Ohmic and super-Ohmic
types at  the Markovian limit \cite{kzc}. Here we shall give the
analytical and exact solution of $Z(\tau)$ for discrete spectra.

 We found that the feature of $Z(\tau)$ depends
on a relationship between two intrinsic energy scales: $M\Omega^2$
and $\sum_{k=1}^N\frac{C_k^2}{m\omega_k^2}$.  The former is the
intrinsic energy scale of the system in terms of the mass $M$ of
the central particle and the corresponding frequency $\Omega$ in
the harmonic trap. The latter is associated with an interaction
energy scale between the system and the environment characterized
by the coupling constants $\{C_k;k=1,\cdots,N\}$ and the intrinsic
energy scales of the environmental oscillators
$\{m\omega_k^2;k=1,\cdots,N\}$. We call $M\Omega^2$ the bounding
strength of the system because it is just the second derivative of
the harmonic potential at its valley. This strength tells how
strongly the particle is bounded in this harmonic trap.  We name
$\sum_{k=1}^N\frac{C_k^2}{m\omega_k^2}$ simply the spectral
strength. These two energy scales indeed act as two time scales in
the total system, one corresponds to the time scale of the system
and the other corresponds to the time scale of the environment
accompanied with the couplings between the system and the
environment. Since the environment contains many different
frequencies, it is not easy to figure out a definite time scale.
However, the spectral strength
$\sum_{k=1}^N\frac{C_k^2}{m\omega_k^2}$ contains the contribution
of every individual frequency in the environment plus the
corresponding coupling to the system. It naturally provides an
alternative expression to the time scale of the environment.

We will see that different relationship between these two
strengthes (or equivalently between two time scales of the system
and environment) alters the overall form of the solution to the
dynamics of the system. The explicit forms of $Z(\tau)$ in cases
that the bounding strength of the system is stronger than, weaker
than or equal to the spectral strength are quite different. We
just present here the results and leave the verification of the
solutions $Z(\tau)$ in Appendix. In general, $Z(\tau)$ has the
following form
\begin{equation}
Z(\tau)=\sum_{k=0}^{N}
\frac{\prod_{j=1}^{N}(\omega_j^2-\nu_{k}^2)} {\prod_{j=0,j\ne
k}^{N}(\nu_{j}^2-\nu_{k}^2)}
\frac{\sin{(\nu_{k}\tau)}}{\nu_{k}}\label{osc}
\end{equation}
where $\nu_n$ satisfy
\begin{equation}
M(\Omega^2-\nu_n^2)\prod_{i=1}^N(\omega_i^2-\nu_n^2)-\sum_{k=1}^N
C_k^2/m\prod_{i\ne k,i=1}^N(\omega_i^2-\nu_n^2)=0 \label{fre1}
\end{equation}
for $n=0,\cdots,N$. When
$M\Omega^2>\sum_{k=1}^N\frac{C_k^2}{m\omega_k^2}$, we have
\begin{equation}
0<\nu_0<\omega_1<\nu_1<\omega_2<\cdots<\nu_{N-1}<\omega_N<\nu_{N}<\infty
\label{fd1}. \end{equation}  For a discrete spectrum, this
constitutes an oscillation function in time.  The continuous limit
will lead this summation over sine functions to a Fourier integral
that may not always be oscillatory.  When
$M\Omega^2=\sum_{k=1}^N\frac{C_k^2}{m\omega_k^2}$, $\nu_0=0$ and the
solution becomes
\begin{eqnarray}
&& Z_=(\tau)=
\Big(\prod_{k=1}^{N}\frac{\omega_{k}^2}{\nu_{k}^2}\Big)\tau
 \nonumber \\ && ~~~~~~ - \sum_{k=1}^{N}\frac{\prod_{j=1}^{N}
(\omega_j^2-\nu_k^2)} {\nu_{k}^2\prod_{j=1,j\ne k}^{N}
(\nu_{j}^2-\nu_{k}^2)}\frac{\sin{(\nu_{k}\tau)}}{\nu_{k}}
\label{lfz}
\end{eqnarray}
with all the other $\nu_n$'s, $N\ge n\ge1$, still being located by
(\ref{fd1}). In the case that
$M\Omega^2<\sum_{k=1}^N\frac{C_k^2}{m\omega_k^2}$, $\nu_0=i\mu_0$
where $\mu_0$ is a real number and $Z(s)$ becomes
\begin{eqnarray}
&& Z_<(\tau)={\prod_{j=1}^{N}(\omega_j^2+\mu_0^2)\over
\prod_{j=1}^{N}(\nu_{j}^2+\mu_0^2)}\frac{\sinh{(\mu_0
\tau)}}{\mu_0}
\nonumber \\
&& ~~~ -\sum_{k=1}^{N}
\frac{\prod_{j=1}^{N}(\omega_j^2-\nu_{k}^2)}
{(\nu_{k}^2+\mu_0^2)\prod_{j=1,j\ne k}^{N}(\nu_{j}^2-\nu_{k}^2)}
\frac{\sin{(\nu_{k}\tau)}}{\nu_{k}}\label{dfz}.
\end{eqnarray}

As we have seen when the bounding strength of the system is
stronger than the spectral strength, the oscillatory nature of the
system is maintained although its oscillation details are
completely altered. Coupling to the $N$ oscillating modes of the
environment generates $N$ new oscillating modes to the system's
dynamics. Combining with the original oscillating mode $\Omega$,
the system has now totally $N+1$ oscillating modes, as shown in
Eq.~(\ref{osc}). Note that the existence of $N+1$ new oscillating
modes was also obtained by Haake and Reinhold \cite{hak} through a
normal mode transformation to the model Hamiltonian but the
physical picture is different. In \cite{hak}, the $N+1$ new
oscillating modes is just a hybridization to the original
frequencies $\{\Omega, \omega_k, k=1, \cdots, N\}$ of the $N+1$
harmonic oscillators for the system plus the environment, due to
the coupling between them. Here the solution of the system, after
the environment's degrees of freedom are integrated out, contains
$N+1$ oscillating modes. It indicates that after the environment
is traced over, the system is no longer a particle trapped in a
harmonic potential, as manifested by the solution of
Eq.~(\ref{osc}). The $N+1$ oscillating modes $\{\nu_n, n=0,
\cdots, N\}$ are determined by Eq.~(\ref{fre1}).

Furthermore, when $M\Omega^2 \rightarrow
\sum_{k=1}^N\frac{C_k^2}{m\omega_k^2}$, we find from
Eq.~(\ref{fre1}) that the frequency $\nu_0 \rightarrow 0$. In
particular, the condition $M\Omega^2 = \sum_{k=1}^N\frac{C_k^2}
{m\omega_k^2}$ leads exactly to $\nu_0 = 0$.  As we see, the first
term in Eq.~(\ref{lfz}) comes from $\nu_0=0$, while the rest $N$
mode oscillations in the solution of $Z(\tau)$ are induced by the
coupling of the system to the $N$ modes of the environment. This
solution indicates that at the critical energy condition,
$M\Omega^2 = \sum_{k=1}^N\frac{C_k^2} {m\omega_k^2}$, the central
particle is driven out of the harmonic potential by its
interaction with the environment. After a relative long time, the
solution of $Z(\tau)$ is dominated by the linear term in time,
which is responsible for an irreversible process as a sign for the
possible rise of forever loss of quantum coherence.

Now, we further consider the case of $M\Omega^2<\sum_{k=1}^N
\frac{C_k^2}{m\omega_k^2}$. It is interesting to see from
Eq.~(\ref{fre1}) that $\nu_0$ becomes an imaginary number.
Although the solution Eq.~(\ref{dfz}) still contains $N$
oscillating modes induced from the coupling of the system to the
$N$ modes in the environment, the imaginary root $\nu_0=i\mu_0$
provides the solution of $Z(\tau)$ with a component that
exponentially grows up in time. The stronger the spectral strength
is, the larger $\mu_0$ will be (see the appendix A.1). In other
words, when the bounding strength of the system, $M\Omega^2$, is
below a critical value of the environment's spectral strength, the
system will be pulled out from the harmonic bounding potential
very quickly and its dynamical process becomes irreversible in an
exponential growth rate.

In fact, the existence of two energy scales in the Hamiltonian
Eq.~(\ref{hamil}) and the corresponding non-Markovian dynamics
resulting from the competition between the two energy scales
 have not be paid attention in the literature. This
is mainly because a counter-term, $\sum_{i=1}^N{C_i^2\over
2m\omega_i^2}X^2$, is usually added to Eq.~(\ref{hamil}) in order
to study the generic behavior of the dissipation dynamics for a
damping harmonic oscillator \cite{clann}, or a "positivity
condition", $\Omega^2-\sum_{i=1}^N {C_i^2 \over \omega_i^2} \geq
0$ (for the case $M=m$), is imposed on the coupling constants and
the unperturbed frequencies to ensure the Hamiltonian having a
finite lower bound \cite{hak}. Obviously the counter-term or the
positivity condition excludes the dynamics corresponding to the
case $M\Omega^2 \leq \sum_{k=1}^N \frac{C_k^2}{m\omega_k^2}$.
However, in the present work we deal with the environment having
only a finite number of discrete modes in which no high frequency
cut-off needs to be introduced. Therefore, no counter-term exists
according to the standard renormalization theory\cite{rent}. Also,
as Caldeira and Leggett had extensively discussed \cite{clann} the
coupling-induced frequency-renormalization effect in many cases is
indeed a real physical effect that has to be taken into account.
There are many physical situations, such as phase transitions of
early universe in extreme environments in cosmology \cite{bran},
fission of a heavy nucleus with coupling of the collective degree
of freedom to the single-particle modes in nuclear physics
\cite{clann}, atomic tunneling with phonon coupling in glasses
\cite{sethna}, and the instability of Bose-Einstein condensation
influenced by the trapping field in cold atoms \cite{santos}, etc.
where the renormalization effect can be very large to reach the
regime $M\Omega^2 \leq \sum_{k=1}^N \frac{C_k^2}{m\omega_k^2}$
such that the system can render the original potential minimum
unstable. The corresponding dynamics will undergo a catastrophic
change that can be observed experimentally. Therefore the
positivity condition should be also not applicable to these
physical situations, in contrast to the study of
environment-induced dissipative or damping harmonic oscillation in
the literature \cite{hak,wss}.

In order to make a comparison between the environments with
discrete and continuous spectra, we take the continuum limit of
the spectrum of the environment. We first define a polynomial
whose roots are those frequencies $\nu_k$, $k=0,\cdots, N$.  This
is the polynomial of Eq.~(\ref{fre1}), $P(\nu)=Mp(\nu)W(\nu)$
where $p(\nu)\equiv\Omega^2-\nu^2-{1\over
M}\sum_{j=1}^N{C_j^2\over m}{1\over\omega_j^2-\nu^2}$ and
$W(\nu)=\prod_{j=1}^N(\omega_j^2-\nu^2)$. We observe that the
general form of $Z(\tau)$ can also be rewritten as
$Z(\tau)=\sum_{k=0}^N{-2\over p'(\nu_k)}\sin(\nu_k \tau)$, where
$p'$ denotes the derivative of $p$ with respect to $\nu$.  By the
fact that the roots of $P(\nu)$ are also that of $p(\nu)$ and the
roots are paired due the evenness of the polynomials, we further
rewrite $Z(\tau)$ replacing $\sin(\nu_k \tau)$ by ${e^{i\nu_k
\tau}-e^{-i\nu_k \tau}\over2i}$ as
$Z(\tau)=-i\sum_{k=0}^{2N+1}{e^{-i\nu_k \tau}\over p'(\nu_k)}$
enumerating
$-\nu_0=\nu_{N+1},-\nu_1=\nu_{N+2},\cdots,-\nu_N=\nu_{N+N+1}$.  By
the residue's theorem, $Z(\tau)$ becomes
${-1\over2\pi}\oint_Cdz{e^{-iz \tau}\over p(z)}$ where the
integration contour encloses all the roots of $p(\nu)$.  Then
taking the continuum limit of $p(\nu)$ that
$\sum_{j=1}^N{C_j^2\over
m}{1\over\omega_j^2-\nu^2}\rightarrow\int_0^\infty
d\omega{D(\omega)C^2(\omega)\over m}{1\over\omega^2-\nu^2}$ where
$D(\omega)$ is the density of states in the environment and
setting ${D(\omega)C^2(\omega)\over
2m\omega}={2M\gamma_0\over\pi}\omega\Theta(\Lambda-\omega)$ as the
spectral density of an Ohmic environment (here $\Theta$ is the
Heavyside function), we reproduce
\begin{equation}
Z(\tau)=e^{-\gamma_0 \tau}{\sin(\tilde{\Omega}
\tau)\over\tilde{\Omega}}\label{contz}
\end{equation}
as a solution to the corresponding version of dissipation
dynamical equation Eq.~(\ref{dissipz}):
$\ddot{Z}(\tau)+2\gamma_0\dot{Z}(\tau)+\Omega_r^2Z(\tau)=0$. Here
$\gamma_0$ is a constant that is usually very small in comparison
to the high frequency cutoff $\Lambda$, and
$\tilde{\Omega}=\sqrt{\Omega_r^2-\gamma_0^2} \simeq \Omega_r$ (for
$\gamma_0 \ll \Lambda$) with the system's renomalized frequency
$\Omega_r^2\equiv\Omega^2-4\gamma_0\Lambda/\pi$ \cite{cal,hu,jpz}.
As we see, whenever the environment's spectrum becomes continuous,
the form of $Z(\tau)$ as a summation over sine functions in
Eq.~(\ref{osc}) becomes a Fourier integral and it may not always
be oscillatory. In other words, the interaction of the system with
a discrete spectral environment leads to very different results
from the system interacting with a continuous spectrum
environment.

Having obtained the analytical solution to the dissipation
dynamical equation for an environment with discrete spectrum and
reproduced the previous known results for the Ohmic bath in the
continuous spectrum limit, we can now utilize this analytical
solution to study the wave packet dynamics in the next section.

\section{Non-Markovian Wave Packet Dynamics}
The geometry of a wave packet is well described by the covariation
matrix \cite{zhg}:
\begin{equation}
\left[
\begin{array}{cc}
   \Delta X^2 & \Delta \{XP\} \\
   \Delta \{PX\} & \Delta P^2 \\
   \end{array} \right] , \nonumber
\end{equation}
where $\Delta X^2$ and $\Delta P^2$ are the widthes of the wave
packet in position and momentum spaces respectively, and are
defined via $\Delta X^2=\left<X^2\right>-\left<X\right>^2$,
$X_c=\left<X\right>$, $\Delta
P^2=\left<P^2\right>-\left<P\right>^2$ and $P_c=\left<P\right>$
with the bracket $\left<.\right>$ denoting the expectation value
of an operator in the state $\rho_A$. $\Delta\{XP\}$ is a quantity
that measures the internal correlation between the position and
 momentum observables, $
\Delta\{XP\}=\frac{1}{2}\left<XP+PX\right>-\left<X\right>
\left<P\right> = \Delta\{PX\}$. The density matrix of the initial
wave packet of the system generally takes the form:
\begin{widetext}
\begin{eqnarray}
\rho_A(R_0,r_0,0) = N_0 e^{ -{1\over 2\Delta X_0^2} \big[(R_0 -{i
\over \hbar}\Delta\{XP\}_0r_0 )^2 + {1\over \hbar^2}\Delta
X^2_0\Delta P^2_0r_0^2 + 2X_0 R_0 +{2i \over
\hbar}(P_0-\Delta\{XP\}_0 X_0 ) r_0 \big]} , \label{rdm0}
\end{eqnarray}
\end{widetext} where $N_0$ is the normalization constant, and the
subscript $"0"$ here denotes the initial values of these quantities.

The density matrix at time $t$ is determined by the time evolution
equation,
\begin{equation} \label{rdm} \rho_{A}(R,r,t)=\int
dR_0dr_0 J(R,r,t|R_0,r_0,0)\rho_{A}(R_0,r_0,0)
\end{equation}
where the propagator $J(R,r,t|R_0,r_0,0)$ (see Eq.~(\ref{prop}))
is easy to obtain once the dynamical equations
(\ref{dissip1}-\ref{dissip2}) are  solved exactly. Explicitly, the
solution of Eq.~(\ref{rdm}) is
\begin{widetext}
\begin{equation} \rho_A(R,r,t) =
\tilde{N} e^{ -{1\over 2\Delta X^2(t)} \big[\big(R -{i \over
\hbar}\Delta\{XP\}(t)r \big)^2 + {1\over \hbar^2}\big(\Delta
X^2(t)\Delta P^2(t)\big)r^2 +2X_c(t)R +{2i \over
\hbar}\big(P_c(t)-\Delta\{XP\}(t)X_c(t) \big) r \big] } \label{rdm2}
\end{equation}
\end{widetext} which keeps the same form as the initial wave packet
(\ref{rdm0}), and $\tilde{N}$ is a normalization factor. The time
evolution of the wave packet is thus completely determined by the
time evolution of the covariation matrix elements, which can be
found from Eqs.~(\ref{prop}) and (\ref{rdm}). Together with the
solutions of $Z(t)$ discussed in the last section, the full
information of the wave packet dynamics, in particular, the
decoherence process can be analyzed now.

The decoherence behavior is mainly characterized by the decay of
the off-diagonal element of the reduced density matrix. In the
literature, to make the quantum coherence of wave packets
manifestation, one usually starts with a superposition of two
well-separated Gaussian wave packets where the decoherence can be
measured directly from the interference of two wave packets
\cite{zuk,braun,str}. However, a single wave packet itself
represents a macroscopic quantum state which has its own physical
interests in many physical systems described by wave packets. A
decoherence measure to a single wave packet is desirable.

Recall that $\rho_A(R,r,t) \equiv \langle x | \rho_A(t) |x'
\rangle$ with $R=(x+x')/2$ and $r=x-x'$, the off-diagonal matrix
element is given by the $r$-dependent part in (\ref{rdm2}) which
describes the quantum coherence dynamics of the wave packet. In
fact, Eq.~(\ref{rdm2}) tells that the dynamics of the off-diagonal
matrix element is fully determined by the covariation matrix
$\Delta X^2(t), \Delta P^2(t), \Delta\{ XP \}(t)$. In particular,
the correlator $\Delta\{XP\}(t)$ describes the phase dynamics of
the density matrix, while the Heisenberg uncertainty, $\Delta
X^2(t) \Delta P^2(t)$, measures the amplitude dynamics of its
off-diagonal behavior. Note that because of its independence on
the widths $\Delta X^2(t)$ and $\Delta P^2(t)$ as well as the
correlator $\Delta\{XP\}(t)$, the center motion of the wave packet
(given by $X_c$ and $P_c$) does not change the geometry of the
wave packet.  Without loss of the generality, we may let the
initial position and the momentum of the wave packet be zero,
$X_0=P_0=0$, then the reduced density matrix simply becomes:
\begin{eqnarray}
&& \rho_A(R,r,t) = \tilde{N} \exp \Big[ -{R^2\over 2 \Delta
X^2(t)}+{i\over \hbar}
{\Delta\{XP\}(t)\over \Delta X^2(t)} Rr  \nonumber \\
&& ~~~~~~~ -\Big({\Delta X^2(t)\Delta P^2(t)-\Delta\{XP\}^2(t)\over
2\hbar^2\Delta X^2(t)}\Big) r^2\Big]. \label{rdm3}
\end{eqnarray}
Now it becomes clear that the decay of the off-diagonal matrix
element is determined by the quantity $D_c(t) \equiv {\Delta
X^2(t)\Delta P^2(t)- \Delta\{XP\}^2(t)\over \Delta X^2(t)}$, where
the denominator $\Delta X^2(t)$ measures the delocalization of the
waver packet influenced by the environment, while the numerator
$\Delta X^2(t)\Delta P^2(t)-\Delta\{XP\}^2(t)$ is actually the
Schr\"{o}dinger uncertainty \cite{sch}. The minimum
Schr\"{o}dinger uncertainty has been used as a criterion to
examine if a wave packet is a squeezed coherent state \cite{zhg}.
Thus $D_c(t)$ is a natural quantity to characterize the
delocalization of a wave packet accompanied with wave packet
decoherence. The degree of quantum decoherence in a wave packet
can then be extracted from the off-diagonal matrix element:
\begin{equation} e^{-{1\over 2\hbar^2}D_c(t) r^2}
= e^{-{1\over 2 \hbar^2}{\Delta X^2(t) \Delta P^2(t)-
\Delta\{XP\}^2(t)\over \Delta X^2(t)} r^2}. \label{chm}
\end{equation}
The larger $D_c(t)$ is, the less there is quantum coherence.

To demonstrate the non-Markovian decoherence dynamics of the wave
packet, we plot the time evolutions of the wave packet width and
the corresponding decoherence measure at different physical
conditions. Note that for an environment with a few modes, the
parameter $\beta$ bares no sense of thermodynamics and it is just
a parameter in the initial state of the environment. We set here
$M=m=\Omega=\hbar=1$, $\beta^{-1}=1.15$ and
$C_k=M\Omega^2\gamma\sqrt{{\Gamma^2\over(\omega_k-\Omega)^2+\Gamma^2}}$
with $\Gamma=500$. The parameter $\gamma$ is dimensionless and is
set differently according to the relationship between $M\Omega^2$
and $\sum_{k=1}^N\frac{C_k^2}{m\omega_k^2}$.  We take an initial
wave packet as $\Delta X_0=1/5$, $\Delta P_0=5$ and
$\Delta\{XP\}_0=X_0=P_0=0$.  We first fix the frequency
distribution in the environment and vary only $\gamma$ to obtain
the conditions $M\Omega^2>\sum_{k=1}^N \frac{C_k^2}{m\omega_k^2}$,
$M\Omega^2=\sum_{k=1}^N \frac{C_k^2}{m\omega_k^2}$ and
$M\Omega^2<\sum_{k=1}^N \frac{C_k^2}{m\omega_k^2}$.  We then fix
$\gamma$ and change the frequency distributions to satisfy the
above three relationships between the bounding and the spectral
strengthes.  The results are shown Figs.~1-2.  From the plots, one
can see that when the bounding strength is larger than the
spectral strength, the wave packet keeps oscillating in its width
as well as the decoherence measure. There is no unidirectional
growth of the decoherence measure. When the two strengthes
balance, the wave packet starts to spread. But the decherence
measure still oscillates at some finite values. The monotonic loss
of quantum coherence is seen when the spectral strength wins over
the bounding strength. This happens also together with an even
faster spreading of the wave packet.  These phenomena are seen no
matter $\gamma$ is varied with the frequency distribution fixed or
the frequency distribution is changed with $\gamma$ fixed (see
Fig.~1, and the solid-curves in Fig.~2). Consequently, the
emergence of decoherence is subjected to the definite greatness of
the spectral strength in comparison to the bounding strength of
the system here.  If the spectral strength is just equal to the
bounding strength, delocalization dynamics occurs without
necessarily leading to decoherence.
\begin{widetext} ~~
\begin{figure}[ht]
\includegraphics[width=14cm, height=10cm]{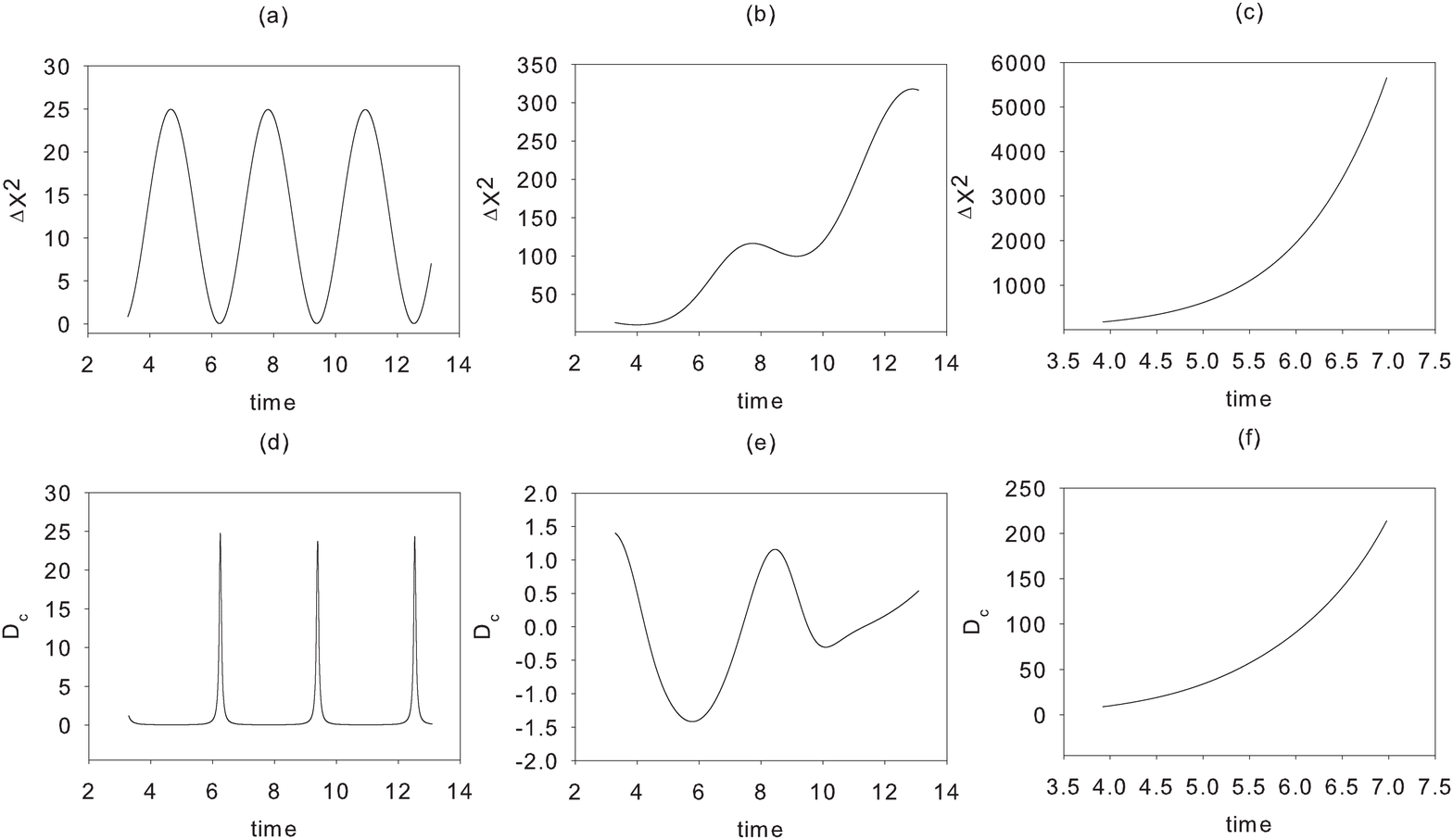}
\caption{(a), (b) and (c) show the time evolution of the width
$\Delta X^2$ of the wave packet at the condition $M\Omega^2>, =$
and $ < \sum_{k=1}^N \frac{C_k^2}{m\omega_k^2}$, respectively.
(d), (e) and (f) the time evolution of the decoherence measure
$D_c$ at the corresponding conditions. The frequencies are chosen
as $\{\omega_1,\omega_2,
\omega_3,\omega_4,\omega_5\}$=$\{0.48,0.86, 1.72,1.84, 1.89 \}$ .
The coupling constant $\gamma$ is adjusted as 0.01, 0.39 and 0.58
to satisfy the bounding strength being larger than, equal to and
smaller than the spectral strength. The wave packet delocalizes
whenever its bounding strength is no larger than the spectral
strength. $D_c$ grows up monotonically when
$M\Omega^2<\sum_{k=1}^N \frac{C_k^2}{m\omega_k^2}$.  The balance
condition $M\Omega^2=\sum_{k=1}^N \frac{C_k^2}{m\omega_k^2}$
results in slow diffusion of the wave packet, slow compared to
that for $M\Omega^2<\sum_{k=1}^N \frac{C_k^2}{m\omega_k^2}$,
without rendering unidirectional loss of quantum coherence.}
\end{figure}
\end{widetext}

The non-Markovian decoherence dynamics of the wave packet (the
memory effect induced from the back-action of the environment) can
be seen by contrasting the exact solution with Markovian
approximation which is physically valid when the time scale of the
environment is much smaller than that of the system and has been
widely used in the literature \cite{wss,breuer}. Mathematically,
the Markovian approximation takes the history independence of the
dynamical variable in Eq.~(\ref{dissipz}), namely $\int_{0}^\tau
d\tau'K_I(\tau-\tau'){Z}(\tau')\simeq Z(\tau)\int_{0}^\tau
d\tau'K_I(\tau-\tau')$ in solving the dissipation dynamical
equation \cite{an07}.   When the time scales of the environments
and the system are comparable (equivalently when the values of the
two strengthes approach to each other), we expect the dynamics of
the wave packet under Markovian approximation deviates obviously
from the exact solution. The three frequency distributions are
chosen such that the time scales of the environments fall upon the
regimes of interests.

As we can see from Fig.~2, when the environment contains only
frequencies much larger than the central one, the system is
located in the regime $M\Omega^2 \gg \sum_{k=1}^N
\frac{C_k^2}{m\omega_k^2}$. We find that the Markovian
approximation well agrees with the exact solution. However, when
the frequencies of the environmental oscillators all lie near the
central frequency (strongest memory regime) where the two
strengthes are very close to each other, an apparent difference
between the Markovian and non-Markovian results shows up. The
exact time evolution of the wave packet width exhibits several
kinks and bumps that are smeared when Markovian processes are
asserted. The extension of wave packet delocalization is over
estimated by Markovian approximation.  The exact time evolution of
the decoherence measure in this time scale shows oscillatory
features that are not seen from the Markovian approximation.
 The magnitude of the decoherence measure is
also over evaluated in the Markovian approximation. In other
words, ignoring the history dependence smears out the subsequent
features of the time evolutions of $\Delta X^2$ and $D_c$ in this
case. When it only consists of lower frequencies, namely the
environment has a longer time scale than that of the system, the
system is easy to fall into the regime $M\Omega^2<\sum_{k=1}^N
\frac{C_k^2}{m\omega_k^2}$ where we find the Markovian
approximation over evaluates the magnitudes of $\Delta X^2$ and
$D_c$ but not as severe as that in the balance case.
\begin{widetext} ~~
\begin{figure}[ht]
\begin{center}
\includegraphics[width=14cm, height=10cm]{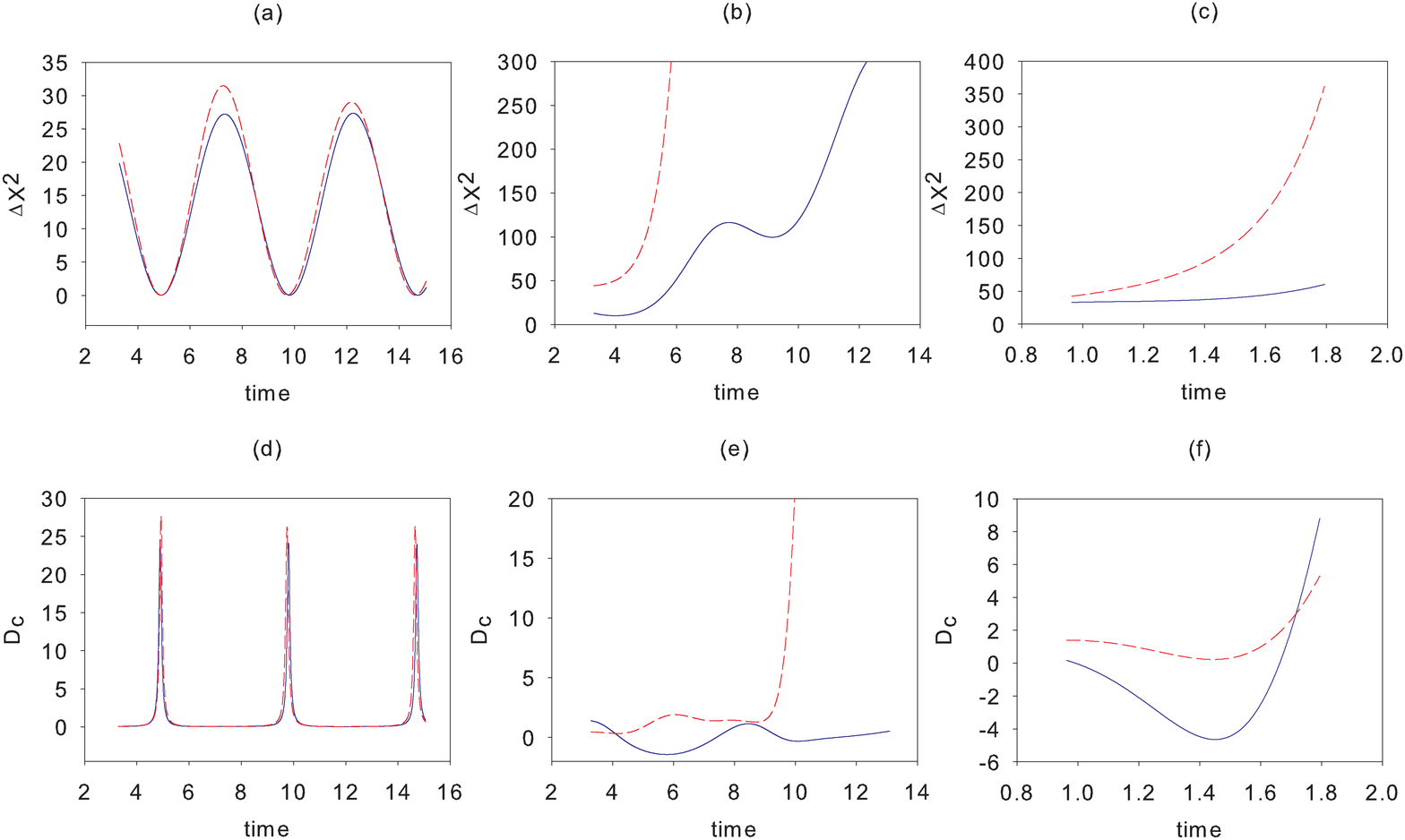}
\caption{The solid blue lines are exact results and the dashed red
lines are obtained with Markovian approximation. (a) and (d) are
for $M\Omega^2>\sum_{k=1}^N \frac{C_k^2}{m\omega_k^2}$.  (b) and
(e) for $M\Omega^2=\sum_{k=1}^N \frac{C_k^2}{m\omega_k^2}$. (c)
and (f) for $M\Omega^2<\sum_{k=1}^N \frac{C_k^2}{m\omega_k^2}$.
The three conditions are achieved by fixing $\gamma$ at 0.39 and
changing the frequency distribution $\{\omega_1,\omega_2,
\omega_3,\omega_4,\omega_5\}$ to be $\{ 2.43, 2.66,2.69,2.70,
2.77\}$ or $\{0.19, 0.23, 0.44, 0.89, 0.96\}$ for $M\Omega^2>$ or
$< \sum_{k=1}^5\frac{C_k^2}{m\omega_k^2}$.  The balance condition
uses the same set of frequencies as in Fig.~1. Markovian
approximation breaks down severely when $M\Omega^2$ is comparable
to $\sum_{k=1}^N \frac{C_k^2}{m\omega_k^2}$.}
\end{center}
\end{figure}
\end{widetext} ~~

The wave packet dynamics we obtained here is not limited to
environments with discrete spectra.  We may consider the wave
packet evolution in an Ohmic bath (a continuous spectrum
environment) as an example. In this case, the system undergoes a
underdamping process according to (\ref{contz}). We find that when
$t\rightarrow\infty$ the width of the wave packet $\Delta X^2(t)$
is given by
\begin{eqnarray}\Delta
X^2(t\rightarrow\infty) = {\hbar\over\pi}\int_0^\infty
d\omega\coth(\beta{\hbar\omega\over2})~~~~~~~~~~~~~~~~\nonumber \\
\times \Big({1\over
M}{\gamma_0\omega\over(\Omega_r^2-\omega^2)^2+4\gamma_0^2
\omega^2}\Big).\label{lmdx} \end{eqnarray} Note that in the linear
response theory, a harmonic oscillator of frequency $\Omega_r$
reacts to an external force with the imaginary part of its
response being $\sigma(\omega)=\Big({1\over
M}{\gamma_0\omega\over(\Omega_r^2-\omega^2)^2+4\gamma_0^2\omega^2}\Big)$.
Eq.~(\ref{lmdx}) can be rewritten as $\Delta
X^2(\infty)={\hbar\over\pi}\int_0^\infty
d\omega\coth(\beta{\hbar\omega\over2})\sigma(\omega)$ which is a
well-known result of the fluctuation-dissipation theorem
\cite{cal,lan}.  In the extremely underdamped limit, the spread of
the wave packet ultimately goes to an equilibrium value: $\Delta
X^2(\infty)={\hbar\over2M\Omega_r}\coth(\beta{\hbar\Omega_r\over2})$,
the same result obtained in the equilibrium quantum statistical
mechanics  \cite{blo,fey,cal}.  The numerical evaluations of
$\Delta X^2(t)$ at large times under such a circumstance also
accurately match the equilibrium quantum statistical result.

\section*{IV CONCLUSIONS AND DISCUSSIONS}
Here we summarize what we have done in this paper.  We study the
dynamics of a particle in a harmonic trap linearly coupled to a
set of noninteracting harmonic oscillators.  We obtained an exact
analytical solution of the system with the environment containing
discrete $N$ modes.  By taking the continuous spectrum limit, we
examined the dynamics in an Ohmic spectral density and reproduced
the solution once obtained by others. In order to visualize the
dynamics more vividly, we let the system start with a wave packet.
The decoherence processes are described in terms of the geometry
of the wave packet. We then closely looked at the wave packet's
localization and delocalization behavior against its decoherence
measure defined by the ratio of Schr\"{o}dinger uncertainty over
the delocalization extension of the wave packet. We observe two
intrinsic energy scales, the bounding strength $M\Omega^2$ and the
spectral strength $\sum_{k=1}^N\frac{C_k^2}{m\omega_k^2}$, acting
as the time scales of the system and the environment,
respectively. These two energy scales are veiled in the previous
investigations when a counter-term was added to the Hamiltonian in
the continuous spectral density where a high-frequency cut-off has
to be introduced or a positivity condition is imposed to the
system. However, the relationship between these two intrinsic
energy scales determines the feature of the delocalization as well
as decoherence dynamics of the wave packet in the cases where the
environment has only a few modes so that no counter-term exists or
no positivity condition can be added when decoherence dynamics is
concerned.

We also numerically demonstrated the time evolution of the wave
packet based on the exact solutions we obtained with an
environment of a few modes satisfying a modified Drude's spectral
density. Our results show that a stronger spectral strength drives
the wave packet out of the harmonic bounding potential as a
consequence of the environment-induced instability, which results
in strong entanglement between the system and the environment and
leads to severe decoherence, while a moderate or weaker spectral
strength oscillates its decoherence measure over time. We also
compared the exact results with Markovian approximation. If
Markovian approximation is done ahead, the relationship between
the two strengthes still plays a determinant role in the
consequent wave packet dynamics. The over all trends of wave
packet dynamics are not altered by Markovian approximation when
the bounding strength is much larger than the spectral strength.
But Markovian approximation under the condition of a larger
spectral strength changes the wave packet dynamics very
drastically from the exact solution.  The critical condition that
the two strengthes balance each other manifests the Non-Markovian
wave packet dynamics most significantly. In conclusion, we have
taken a few factors to explore the mechanisms of decoherence of a
wave packet in a few-mode environment. One may attempt to see
these issues in relevant systems mentioned above.

\begin{acknowledgments}
We would like to thank Drs. M. T.~Lee and J. H.~An for fruitful
discussions.  We would also like to specially thank Prof. A. J.
Leggett for his help in clarifying the counter-term problem in
Caldeira-Leggett model. This work is supported by the National
Science Council of Republic of China under Contract
No.~NSC-96-2112-M-006-011-MY3 and No.~NSC-95-2112-M-006-001.
\end{acknowledgments}
\appendix

\section*{APPENDIX: The solutions to the
dissipation dynamical equations}

\subsection*{A.1 Root Property} Before we proceed to justify our
solution to the dissipation dynamical equation, we first look at
(\ref{fre1}).  The roots of this polynomial determine the properties
of $Z(s)$.  We denote the general argument of this polynomial by
$\nu$ instead of $\nu_n$.  For convenience of reference, we write
this polynomial again,
\begin{align}
&P_0(\nu)=\nonumber\\
&M(\Omega^2-\nu^2)\prod_{i=1}^N(\omega_i^2-\nu^2) -\sum_{k=1}^N
C_k^2/m\prod_{i\ne k,i=1}^N(\omega_i^2-\nu^2)\label{poly}.
\end{align}
It is well known that if the polynomial is evaluated at some point
on the real axis to be positive  and is negative at some other
point, then there is a point in between at which this polynomial is
evaluated to be zero as its root, denoted by $\nu_i$.  We arrange
$\omega_1$ through $\omega_N$ in a way that
$\omega_1<\omega_2<\cdots<\omega_N$ and let $\nu=\omega_n$, then
Eq.~(\ref{poly} becomes
\begin{eqnarray}
P_0(\omega_n) = (-1)^n{C_n^2\over
m}\prod_{i=1}^{n-1}(\omega_n^2-\omega_i^2)
\prod_{i=n+1}^N(\omega_i^2-\omega_n^2) .
\end{eqnarray}
Clearly the sign of the polynomial at $\omega_n$ for
$n=1,2,\cdots,N$ is determined by $(-1)^n$, so the roots are located
in the vicinity between adjacent frequencies $\omega_n$ and
$\omega_{n+1}$ and we have located $N-1$ distinct roots that are all
positive.  Since the polynomial is even in its argument, the roots
are paired.  This gives another set of $N-1$ roots that are all
negative.  We can also see there is a root larger than $\omega_N$ by
letting $\omega_i=\alpha_i\Omega$ for $i=1,\cdots,N$ and
$\nu=(\alpha_N+f)\Omega$, that is,
\begin{widetext}
\begin{eqnarray}
P_0((\alpha_N+f)\Omega) &=&
(-1)^{N+1}\prod_{j=1}^N((\alpha_N+f)^2-\alpha_j^2)((\alpha_N+f)^2-1)
\Omega^2  \Big( M\Omega^2-\frac{\sum_{k=1}^N\frac{C_k^2}{m}
\frac{1}{(\alpha_k\Omega)^2+(f\Omega)^2}}{(\alpha_N+f)^2-1} \Big).
\end{eqnarray}
\end{widetext} We can make $P_0(\alpha_N\Omega=\omega_N)$ and
$P_0((\alpha_N+f)\Omega)$ differ by a sign by choosing an $f$ which
is large enough that,
\[((\alpha_N+f)^2-1) \Big(M\Omega^2-\frac{\sum_{k=1}^N\frac{C_k^2}{m}
\frac{1}{(\alpha_k\Omega)^2+(f\Omega)^2}}{(\alpha_N+f)^2-1} \Big)>0
,
\]
 and the sign of $P_0((\alpha_N+f)\Omega)$ is
$(-1)^{N+1}$.  So far by including the negative partner of this
root, we have found $2N$ distinct roots.  We know this polynomial
has $2N+2$ roots so there is still a pair of roots not being located
yet.  Since the sign of the polynomial at $\omega_1$ is minus, if
$P_0(0)$ is positive, then there is a root lying between $0$ and
$\omega_1$.  We see $P_0(0)$ is just the difference between the
bounding strength of the system and the spectral strength times a
positive number.
\begin{equation}
P_0(0)=\left(\prod_{i=1}^{N}\omega_i^2\right)
\left(M\Omega^2-\sum_{k=1}^N{C_k^2\over m\omega_k^2}\right) .
\end{equation}
If $M\Omega^2$ is larger than $\sum_{k=1}^N{C_k^2\over
m\omega_k^2}$, then $P_0(0)$ is positive and we have a root that is
larger than $0$ and smaller than $\omega_1$.  Since $P_0(0)$ is
proportional to the product of all the roots it has, $P_0(0)=0$
implies there is a pair of roots doubly located at the origin.  If
$P_0(0)$ is negative, then we have to look for roots somewhere out
of the real axis.  In this case we define $P_y(y)\equiv P_0(iy)$
where $y$ is in the real domain, then $P_y(0)=P_0(0)<0$.  If we can
find a $y$ such that $P_y(y)>0$, then there is a root between $0$
and $y$ for $P_y$ which implies a root for $P_0$.  Let $y=f\Omega$
where $f$ is a positive number and we have
\begin{eqnarray}
&& P_y(f\Omega)= \prod_{j=1}^{N}\big[(\Omega\alpha_j)^2
+(f\Omega)^2)\big](f^2+1)\nonumber \\
&& ~~~ \times \Big(M\Omega^2-\frac{1}{f^2+1}\sum_{k=1}^{N}
\frac{C_k^2}{m}\frac{1}{(\Omega\alpha_k)^2+(f\Omega)^2}\Big).
\end{eqnarray}
Again, $f$ can be chosen to be large enough such that
\[
\Big(M\Omega^2-\frac{1}{f^2+1}\sum_{k=1}^{N}
\frac{C_k^2}{m}\frac{1}{(\Omega\alpha_k)^2+(f\Omega)^2}\Big)>0 .
\]
This gives us a real root to $P_y$ and also an imaginary root to
$P_0$.  This root of course has a negative partner and we find
finally all the roots.

We summarize the location of half of the roots as
 \begin{equation}
0\leq\nu_0<\omega_1<\nu_1<\omega_2<\cdots<\nu_{N-1}<\omega_N<\nu_{N}<\infty,
\end{equation} for $M\Omega^2\geq\sum_{k=1}^N{C_k^2\over m\omega_k^2}$ and
$\nu_0=i\mu_0$ where $\mu_0$ is a real number for
$M\Omega^2<\sum_{k=1}^N{C_k^2\over m\omega_k^2}$.  The other half is
negatively paired with this.

\subsection*{A.2 Verification of the Solution}

We have proved in the last subsection that $\nu_0$ through $\nu_N$
are $N+1$ distinct numbers.  This ensures that the amplitudes of
each component in the solution of $Z(s)$ do not diverge.  Now we
shall check whether (\ref{osc}) fits (\ref{dissipz}) with the
boundary conditions that $Z(0)=0$ and $\dot{Z}(0)=1$. Substituting
(\ref{osc}) to (\ref{dissipz}), we have
\begin{widetext}
\begin{eqnarray}
(\ddot{Z}(s)+\Omega^2Z(s))+{2\over
M}\int_{0}^sds'K_I(s-s'){Z}(s')&=&
\sum_{k=0}^NA_k\left[(\Omega^2-\nu_k^2)\sin(\nu_k s)-{1\over
M}\sum_{j=1}^N{C_j^2\over m\omega_j}{\omega_j\sin(\nu_ks)
-\nu_k\sin(\omega_js)\over\omega_j^2-\nu_k^2}\right]\nonumber \\
&=& \sum_{j=1}^N{C_j^2\over
m\omega_j}\sin(\omega_js)\sum_{k=0}^N{\prod_{i=1,i\ne
j}^N(\omega_i^2-\nu_k^2)\over\prod_{l=0,l\ne k}^N(\nu_l^2-\nu_k^2)}
, \label{pf}
\end{eqnarray}
\end{widetext}
where $A_k$'s are the amplitudes
$\frac{\prod_{j=1}^{N}(\omega_j^2-\nu_{k}^2)} {\prod_{j=0,j\ne
k}^{N}(\nu_{j}^2-\nu_{k}^2)}{1\over\nu_k}$.  We now introduce an
identity,
\begin{equation}
\sum_{k=1}^{\bar{n}}\frac{b_k^j}{\prod_{i\ne
k,i=1}^{\bar{n}}(b_k-b_i)}=\Bigg\{
\begin{array}{ll}
0, &  1\leq j\leq\bar{n}-2
\\ 1, & j=\bar{n}-1
\\ \sum_kb_k, & j=\bar{n} .
\end{array} \label{idd}
\end{equation}
Comparing the factor $\sum_{k=0}^N{\prod_{i=1,i\ne
j}^N(\omega_i^2-\nu_k^2)\over\prod_{l=0,l\ne k}^N(\nu_l^2-\nu_k^2)}$
in the last line of (\ref{pf}) with the identity (\ref{idd}) by
replacing $b_i$ by $\nu_{i+1}^2$ and $\bar{n}$ by $N+1$, we can see
in the numerator that the highest power of $\nu_k^2$ is $N-1$ and
$N-1=(N+1)-2=\bar{n}-2$.  So this summation goes to zero.  And
$Z(s)$ with the proposed form (\ref{osc}) fits (\ref{dissipz}). The
boundary condition $Z(0)=0$ is easy to check for sine functions are
zero when their arguments are zero. To check the boundary value
satisfied by its first derivative, we take,
\begin{equation}
\dot{Z}(0)=A_k\nu_k=\sum_{k=0}^N{\prod_{i=1}^N(\omega_i^2
-\nu_k^2)\over\prod_{l=0,l\ne k}^N(\nu_l^2-\nu_k^2)}.
\end{equation}
Making use of (\ref{idd}) for the highest power in the numerator now
equaling to $N=\bar{n}-1$, $\dot{Z}(0)$ becomes unity.


\begin{thebibliography}{99}

\bibitem{sch26} E. Schrodinger, Naturwiss. 14, 664 (1926).

\bibitem{zuk} W. H. Zurek, Phys. Today {\bf 44} (10), 36 (1991); Rev. Mod. Phys. \textbf{75},
715 (2003).

\bibitem{fey} R. P. Feynman and F. L. Vernon, Ann.
Phys. \textbf{24}, 118 (1963).

\bibitem{cal} A. O. Caldeira and A. J. Leggett, Physica \textbf{121A}, 587
(1983).

\bibitem{zuk1} W. H. Zurek, Phys. Rev. D \textbf{26}, 1862 (1982).

\bibitem{clann} A. O. Caldeira and A. J. Leggett, Ann. Phys.
\textbf{149}, 374 (1983); \textbf{153}, 445 (1984).

\bibitem{hak}F. Haake, and R. Reibold, Phys. Rev. A \textbf{32},
2462 (1985).

\bibitem{fbq}H. Grabert, P. Schramm, G. -L. Ingold, Phys. Rev.
Lett. \textbf{58}, 1285(1987).

\bibitem{zur} W. G. Unruh and W. H. Zurek, Phys. Rev. D \textbf{40}, 1071
(1989).

\bibitem{hu} B. L. Hu, J. P. Paz, and Y. H. Zhang,
Phys. Rev. D \textbf{45}, 2843 (1992); {\it ibid.} {\bf 47} 1576
(1993).

\bibitem{jpz} J. P. Paz, S. Habib and W. H. Zurek, Phys. Rev. D \textbf{47}, 488
(1993); J. R. Anglin, J. P. Paz and W. H. Zurek, Phys. Rev. A {\bf
55} 4041 (1997).

\bibitem{braun}D. Braun, P. A. Braun, and F. Haake, Opt. Commun. {\bf 179},
411 (2000).

\bibitem{ez}G. W. Ford  and R. F. O'Connell, Phys. Rev. D \textbf{64}, 105020 (2001).


\bibitem{str}W. T. Strunz and F. Haake, Phys. Rev. A {\bf 67},
022102 (2003); W. T. Strunz, F. Haake and D. Braun, {\it ibid.}
{\bf 67}, 022101 (2003).


\bibitem{kazu} K. Shiokawa and B. L. Hu, Phys. Rev. A \textbf{70}, 062106
(2004).

\bibitem{aw}G. W. Ford and R. F. O'Connell, Phys. Rev. A \textbf{73}, 032103 (2006).

\bibitem{an}J. H. An and W. M. Zhang, Phys. Rev. A {\bf 76},
042127 (2007).

\bibitem{Chou07} C. H. Chou, T. Yu, and B. L. Hu, Phys. Rev. E {\bf 77}, 011112 (2008).



\bibitem{brf}C. Brif, H. Rabitz, A. Wallentowitz, I. A. Walmsley,
Phys. Rev. A \textbf{63}, 063404 (2001)


\bibitem{chc}M. Spanner, E. A. Shapiro and M. Ivanov, Phys. Rev.
Lett. \textbf{92}, 093001 (2004); E. A. Shapiro, I. A. Walmsley,
and M. Y. Ivanov, {\it ibid.} \textbf{98}, 050501 (2007)



\bibitem{wong}V. Wong, V. and M. Gruebele, J. Phys. Chem.
\textbf{103}, 10083 (1999); Chem. Phys. \textbf{284}, 29 (2002).

\bibitem{et} A. Damjanovic, I. Kosztin, U. Kleinekathofer and K. Schulten
Phys. Rev. E {\bf 65}, 031919 (2001).





\bibitem{legt} A. J. Legget, S. Chakravarty, A. T. Dorsey, M. P. A.
Fisher, A. Garg and W. Zwerger, Rev. Mod. Phys. \textbf{59}, 1
(1987).

\bibitem{gr}H. Grabert, P. Schramm, and G.-L. Ingold, Phys. Rep. {\bf 168}, 115
(1988).

\bibitem{crm}H. Carmichael, \textit{An open systems approach to quantum
optics} (Springer-Verlag, New York, 1993).

\bibitem{wss} U. Weiss, \textit{Quantum Dissipative Systems} (World Scientific, Singapore,
1999).

\bibitem{breuer} H.-P. Breuer and F. Petruccione, \textit{The
Theory of Open Quantum Systems} (Oxford University Press, Oxford,
New York, 2002).

\bibitem{jh}J. J. Halliwell and T. Yu, Phys. Rev. D \textbf{53},
2012 (1996).

\bibitem{mdru}K. Lindenberg and B. J. West, Phys. Rev. A\textbf{30},
568 (1984); H. Callen and T. Weldon, Phys. Rev. \textbf{83}, 34
(1951); R. Kubo, \textit{Lectures in Theoretical Physics} Vol. 1
(Interscience, New York, 1959), pp. 120-203.

\bibitem{rent} M. Peskin, and D. Schroeder, {\it An Introduction to
Quantum Field Theory}, (Perseus Books Pub., 1995); J. Zinn-Justin,
{\it Quantum Field Theory and Critical Phenomena}, (Oxford
University Press. Oxford, 4th ed. 2002))

\bibitem{clann1} See the discussion in Ref.~[6] on page 388-391,
also in private communications with A. J. Leggett.

\bibitem{fey1} R. Feynman and A. Hibbs, \textit{Quantum Mechanics and Path
Integrals} (McGraw-Hill, New York, 1965).

\bibitem{kzc} K. Shiokawa and R. Kapral, J. Chem. Phys. \textbf{117}, 7852
(2002).

\bibitem{zhg} W. M. Zhang, D. H. Feng and R. Gilmore,
Rev. Mod. Phys. \textbf{62}, 867 (1990).

\bibitem{sch} E. Schodinger, Ber. Kgl. Akad. Wiss. Berlin, 296 (1930).

\bibitem{an07}J. H. An, M. Feng and W. M. Zhang, arXiv: 0705.2472
(2007).

\bibitem{bran}R. H. Brandenberger, Rev. Mod. Phys. {\bf 57}, 1
(1985); A. Linde, {\it Particle Physics and Inflationary
Cosmology} (Harwood Academic, Chur, 1990) and references therein.

\bibitem{sethna}J. P. Sethna, Phys. Rev. B {\bf 24}, 698 (1981).

\bibitem{santos} L. Santos, G. V. Shlyapnikov, P. Zoller, and M.
Lewenstein, Phys. Rev. Lett. {\bf 85}, 1791 (2000).


\bibitem{lan} L. D. Landau and E. M. Lifschitz, \textit{Statistical
Physics} (Pergamon, London, 1969).

\bibitem{blo} F. Bloch, Z. Phys. \textbf{74}, 295 (1932).












\end{thebibliography}
\end{document}